\documentclass[11pt]{article}

\usepackage[margin=0.88in]{geometry}
\usepackage{times}
\usepackage{microtype}
\usepackage{booktabs}
\usepackage{float}
\usepackage{tabularx}
\usepackage{array}
\usepackage{multirow}
\usepackage{enumitem}
\usepackage{amsmath,amssymb}
\usepackage{graphicx}
\usepackage{xcolor}
\usepackage{url}
\usepackage{hyperref}
\usepackage{natbib}
\usepackage{tikz}
\usetikzlibrary{arrows.meta,positioning,fit,shapes.geometric}
\hypersetup{colorlinks=true,linkcolor=black,citecolor=black,urlcolor=blue}
\setlist{nosep,leftmargin=*}

\title{\textbf{Toward Sustainable Distributed LLM Inference:}\\
A Systems Synthesis and Research Agenda for an Energy-, Carbon-, and Cache-Aware \texttt{llm-d} Control Plane}

\author{Twinkll Sisodia\\
\texttt{twinklls@bu.edu}}
\date{September 2026}

\begin{document}
\maketitle

\begin{abstract}
Large language model (LLM) sustainability is increasingly a serving-systems problem, not only a training problem. In production, energy and carbon impact depend on more than model size: workload shape, batching, key--value (KV) cache reuse, prefill/decode placement, model and accelerator choice, power state, geographic carbon intensity, and service-level objectives (SLOs) all matter. Recent systems papers study many of these factors separately. This paper connects those results and asks a practical engineering question: what do they imply when the decision point is a distributed inference control plane such as \texttt{llm-d}?

The contribution here is synthesis, not a new set of benchmark results. Reported performance, energy, carbon, and cost improvements remain the results of the cited papers and systems. I group the literature into recurring design patterns and use those patterns to sketch a \emph{Sustainable Inference Control Plane} (SICP) for \texttt{llm-d}. The proposed control plane would consider latency, energy, carbon, cache reuse, serving cost, and quality when routing and scaling, while keeping TTFT/TPOT SLOs as hard constraints. I also outline an evaluation framework based on SLO-satisfied goodput per joule and per gram CO$_2$e, together with a reproducible experimental plan. The main observation from connecting the literature is that sustainable LLM inference is unlikely to come from one ``green'' model or one accelerator; it is more naturally treated as a control problem across model, phase, cache, hardware, replica, region, and time.
\end{abstract}


\section{Introduction}
The sustainability discussion around modern AI has historically emphasized the large, visible cost of frontier-model training. That framing is becoming incomplete. Production LLMs execute continuously, often across millions or billions of requests, and inference workloads are increasingly lengthened by reasoning, retrieval, tool use, agents, and multimodal generation. Recent empirical work shows that inference energy depends strongly on workload geometry, software stack, accelerator, batching, decoding strategy, and parallelism; naive estimates based on FLOPs or theoretical utilization can substantially mischaracterize real consumption \citep{fernandez2025energy}. Industrial systems research similarly shows that energy, power, carbon, and cost can be reduced while meeting latency SLOs when serving infrastructure dynamically reconfigures itself \citep{stojkovic2025dynamollm,li2025ecoserve,stojkovic2025tapas}.

In parallel, LLM serving has undergone a systems transition. PagedAttention and vLLM reframed KV cache as a virtual-memory problem \citep{kwon2023pagedattention}; Splitwise and DistServe separated prefill from decode because the phases stress hardware differently \citep{patel2024splitwise,zhong2024distserve}; QServe demonstrated model--system co-design through aggressive weight, activation, and KV quantization \citep{lin2024qserve}; Mooncake made distributed KV state a first-class architectural resource \citep{qin2024mooncake}; and newer work extends serving to agentic continuity and composite multimodal graphs \citep{li2025continuum,jha2026mstar}. The practical implication is important: inference is no longer a homogeneous function call behind a round-robin load balancer.

The open-source \texttt{llm-d} project makes that transition explicit. It has evolved from a Kubernetes-native distributed serving stack into an inference control plane that performs cache-aware and load-aware routing, prefill/decode disaggregation, multi-tier KV management, autoscaling, heterogeneous-engine orchestration, predicted-latency scheduling, and request-level tracing \citep{llmd2026v08,llmd2026latency,llmd2026p2p,llmd2026trace,llmd2026v09}. These capabilities expose a natural insertion point for sustainability: the same control plane that asks ``which endpoint will be fastest?'' can also ask ``which feasible endpoint will satisfy the SLO with the least incremental energy, carbon, power pressure, and recomputation?''

The goal of this paper is narrower than designing a new serving engine. I am connecting results that are often discussed separately and asking what they imply at the control-plane level. No experimental gain reported in this paper is an original measurement. Whenever a numerical result or an existing system capability is discussed, it is attributed to the cited work. The SICP architecture, taxonomy, and evaluation plan are the synthesis developed in this paper.

\paragraph{What this paper adds.}
\begin{enumerate}
    \item A common taxonomy that places recent LLM inference and sustainable-computing work across request, model, phase, cache, hardware, cluster, datacenter, and lifecycle layers.
    \item A set of recurring engineering patterns---phase asymmetry, workload sensitivity, state reuse, heterogeneity, SLO-constrained adaptation, and telemetry-driven control---that appear across otherwise separate research lines.
    \item A concrete SICP design sketch for \texttt{llm-d} that connects those patterns to routing, KV placement, power management, model selection, autoscaling, and regional placement.
    \item A measurement and evaluation plan for testing the proposal without treating cost, average power, or a single workload as a proxy for sustainability.
\end{enumerate}

\section{Scope and Evidence Selection}
I used a structured but pragmatic literature selection covering work published or publicly released from 2023 through September 3, 2026. The review emphasizes peer-reviewed papers and arXiv preprints from university systems/AI groups and major industrial research organizations, supplemented by official \texttt{llm-d} architecture and release material only when describing current \texttt{llm-d} capabilities.

The selection focuses on work that directly studies at least one of the following: inference energy or carbon; power/thermal management; prefill/decode disaggregation; KV cache management; scheduling/routing; quantization; speculative decoding; model cascades; heterogeneous accelerators; agentic/multimodal serving; or geographically/temporally carbon-aware inference. Training-only sustainability work is used only as background and is not treated as evidence for serving-system behavior.

This is not a PRISMA-style systematic review and does not claim exhaustive coverage of every paper released during the period. The intent is to collect directly relevant systems results, compare the control variables they expose, and look for patterns that can be connected at the inference-control-plane layer.

\paragraph{Claim and attribution boundary.}
The paper separates prior results from synthesis in three ways. First, empirical findings and numerical improvements are stated with citations to the original papers or project material. Second, descriptions of existing \texttt{llm-d} behavior are limited to cited project documentation. Third, the SICP equations, taxonomy, design choices, research questions, and experimental roadmap are proposals derived from connecting those sources; they are not presented as implemented or experimentally validated results.

\section{What the Recent Literature Converges On}

\subsection{Inference energy is workload-dependent, not a model constant}
A central finding in recent empirical work is that ``energy per query'' is not a stable property of a model. \citet{fernandez2025energy} measure energy across frameworks, decoding strategies, GPU architectures, sequence lengths, batching modes, and parallelism, finding that optimized serving can reduce energy substantially relative to unoptimized baselines, while the best optimization changes with workload geometry. Their measurements also show that decode often dominates energy for generation-heavy workloads, speculative decoding can save energy at small batches but increase it at high batches, and adding tensor-parallel GPUs can reduce latency while increasing total energy.

Taken together, these measurements make static rules---for example, always selecting the newest GPU, always maximizing parallelism, or always enabling speculative decoding---hard to justify. The engineering implication I draw is that a sustainability-aware scheduler should condition its decision on prompt length, expected output length, batch state, accelerator, cache hit, and SLO.

\subsection{Prefill and decode should be treated as different resources}
Splitwise established that prompt computation is compute-intensive while token generation is more memory-intensive and can underutilize the compute capability of high-end GPUs \citep{patel2024splitwise}. DistServe showed that separating the phases can improve SLO-constrained goodput by eliminating phase interference and independently choosing resource allocation and parallelism \citep{zhong2024distserve}. Sarathi-Serve explored the complementary path of chunked prefill and stall-free scheduling to reduce the throughput--latency conflict without full disaggregation \citep{agrawal2024sarathi}. Later work such as TaiChi argues that aggregation and disaggregation are not universally superior; the optimal mode depends on TTFT and TPOT SLO geometry \citep{wang2025taichi}.

Taken together, these papers suggest a sustainability implication beyond performance: phases with different compute and memory behavior can also have different energy-efficient operating points. PowerSlider explicitly studies this asymmetry under time-varying power caps \citep{li2026powerslider}. This motivates testing phase-specific power control rather than assuming one power state is best for an entire model replica.

\subsection{KV state is both a performance asset and an energy asset}
PagedAttention showed that improved KV memory management can unlock much larger effective batches and throughput \citep{kwon2023pagedattention}. Mooncake extended the idea to a distributed KVCache-centric architecture with CPU, DRAM, and SSD resources participating in cache storage and movement \citep{qin2024mooncake}. For agentic applications, Continuum shows that interleaved tool calls create a new decision: retain KV state during a pause or evict and later recompute/reload it \citep{li2025continuum}. Current \texttt{llm-d} work further adds cross-endpoint peer-to-peer KV sharing so an idle endpoint can pull cached state instead of recomputing a long prefix \citep{llmd2026p2p}.

Recomputation is therefore not only a latency penalty; it is avoidable energy. But cache retention and transfer are not free. A sustainable system needs an explicit \emph{transfer-versus-recompute} model that prices network energy, CPU/DRAM/storage occupancy, HBM pressure, and the probability of reuse.

\subsection{Efficiency is a model--software--hardware co-design problem}
QServe demonstrates that lower precision alone is insufficient; performance depends on how quantization maps to GPU kernels and memory behavior \citep{lin2024qserve}. CMU's energy study similarly finds that framework and kernel choices can materially change real energy use \citep{fernandez2025energy}. Heterogeneous serving work such as Splitwise shows that a cheaper/lower-power device can be preferable for decode even if it is not the fastest general-purpose accelerator \citep{patel2024splitwise}. This motivates a sustainability policy that chooses hardware by \emph{phase and workload}, not by a single benchmark score.

\subsection{Quality-aware model routing is a sustainability lever}
FrugalGPT, Hybrid LLM, and RouteLLM show that queries can be routed between models of different capability/cost while maintaining a target quality level \citep{chen2023frugalgpt,ding2024hybrid,ong2024routellm}. Google Research's speculative-cascade direction combines smaller and larger models to reduce unnecessary expensive inference \citep{google2025speccascades}. These works primarily optimize cost or latency, but their structure is directly applicable to sustainability: if a smaller model can satisfy the quality target, running a larger model is often wasted computation. However, cost is not a perfect proxy for energy or carbon, so future routers should learn quality and sustainability jointly.

\subsection{Carbon, power, and embodied impact require control above the model server}
DynamoLLM dynamically adjusts cluster configuration under latency SLOs and reports substantial reductions in energy, operational carbon, and cost \citep{stojkovic2025dynamollm}. EcoServe broadens the accounting boundary to embodied carbon and proposes ``Reduce, Reuse, Rightsize, Recycle'' as design pillars \citep{li2025ecoserve}. TAPAS incorporates power and thermal constraints into placement, routing, and reconfiguration \citep{stojkovic2025tapas}. XWind explores cross-site inference near renewable generation \citep{reddy2026xwind}, while recent live work shows that real-time grid carbon signals can be used as an overlay on production routing \citep{bernhard2026carbonrouting}.

Viewed together, these systems suggest that some sustainability decisions necessarily sit above a CUDA kernel or a single model-server process. Replica-, cluster-, and region-level choices need a higher-level control point.

\begin{table}[H]
\centering
\small
\caption{Representative evidence and the control-plane implication derived from it. Reported numerical gains belong to the cited systems and are not results of this paper.}
\label{tab:evidence}
\begin{tabularx}{\textwidth}{p{2.8cm}p{2.3cm}X X}
\toprule
\textbf{Work} & \textbf{Primary lever} & \textbf{Representative finding} & \textbf{Implication for sustainable distributed inference} \\
\midrule
vLLM / PagedAttention \citep{kwon2023pagedattention} & KV memory & 2--4$\times$ throughput over prior serving baselines at similar latency in reported experiments & Memory fragmentation and cache duplication are sustainability issues because they reduce useful work per accelerator-hour. \\
Splitwise \citep{patel2024splitwise} & P/D phase split & Up to 1.4$\times$ throughput at 20\% lower cost; alternative design points increase throughput under fixed power/cost & Match accelerator class and power state to phase behavior. \\
DistServe \citep{zhong2024distserve} & P/D disaggregation & Large goodput gains under TTFT/TPOT constraints in reported workloads & Optimize sustainability subject to phase-specific SLOs, not average latency. \\
QServe \citep{lin2024qserve} & W4A8KV4 co-design & Higher serving throughput and lower dollar cost than compared baselines & Precision must be selected with hardware/kernel awareness; KV precision is a first-class lever. \\
Energy Considerations \citep{fernandez2025energy} & Workload-aware optimization & Up to 73\% lower energy than an unoptimized baseline; optimization efficacy varies by workload/hardware & Sustainability policies must be conditional and learned from telemetry. \\
DynamoLLM \citep{stojkovic2025dynamollm} & Dynamic cluster management & Reports 53\% energy and 38\% operational-carbon reduction while meeting SLOs & Energy/carbon belong in the cluster controller's objective. \\
EcoServe \citep{li2025ecoserve} & Carbon lifecycle & Reports up to 47\% carbon reduction and highlights embodied-carbon role of host infrastructure & Include lifecycle-aware rightsizing, reuse, and batch deferral. \\
Continuum \citep{li2025continuum} & Agentic KV TTL & Large job-completion gains by retaining KV across tool calls when beneficial & Predict future reuse; price retain/offload/reload/recompute actions. \\
M* \citep{jha2026mstar} & Composite multimodal serving & Dataflow-graph serving improves throughput/latency across composite models & Sustainability control must generalize beyond text autoregressive loops. \\
PowerSlider \citep{li2026powerslider} & Phase-aware power control & Maintains substantially more goodput under changing power caps than reported baselines & Couple routing/disaggregation with per-phase DVFS and grid demand response. \\
llm-d predicted-latency routing \citep{llmd2026latency} & Learned routing & Online model uses prompt length, cache hit, queue depth, KV pressure, etc. to predict TTFT/TPOT & Extend the predictor from latency-only to multi-objective latency+energy+carbon. \\
\bottomrule
\end{tabularx}
\end{table}

\section{A Taxonomy of Sustainable Inference Levers}
For this synthesis, I use eight control layers. The boundaries are not strict; the point is to make the interactions visible, because optimizing one layer in isolation can worsen another.

\begin{enumerate}
    \item \textbf{Request layer:} prompt length, output budget, reasoning mode, deadlines, priority, batchability, and expected reuse.
    \item \textbf{Model layer:} model size, dense versus MoE, quantization, speculative-draft configuration, LoRA/adapters, and model cascade choice.
    \item \textbf{Phase/operator layer:} prefill, reasoning/thinking, answer decode, encoders, diffusion heads, experts, and other composite operators.
    \item \textbf{State layer:} KV cache locality, precision, TTL, offload tier, replication, peer transfer, and recomputation.
    \item \textbf{Hardware layer:} accelerator family, memory capacity/bandwidth, frequency/power state, tensor/expert/pipeline parallelism, and interconnect.
    \item \textbf{Replica/cluster layer:} queue depth, batch formation, routing, flow control, autoscaling, consolidation, and failure recovery.
    \item \textbf{Datacenter/region layer:} PUE, thermal headroom, power caps, electricity price, carbon intensity, water stress, and renewable availability.
    \item \textbf{Lifecycle layer:} embodied carbon, hardware lifetime, host overprovisioning, reuse, and recycling.
\end{enumerate}

A sustainable control plane should not necessarily optimize all eight layers on every request. Instead, it should expose a hierarchy: fast millisecond-scale routing decisions, slower seconds/minutes autoscaling and power control, and still slower regional capacity/lifecycle decisions.

\section{Why \texttt{llm-d} Is a Natural Sustainability Control Point}
\texttt{llm-d} is designed above the model-serving engine, allowing vLLM, SGLang, and other backends to focus on efficient execution while the control plane manages distributed placement and orchestration. By v0.8, the project explicitly described itself as an inference control plane and expanded beyond a single engine and a single text-serving pattern \citep{llmd2026v08}. By v0.9, high availability, GPU-utilization-aware routing, hardened flow control, broader autoscaling foundations, and support for additional workload classes had moved the architecture further toward a general production control layer \citep{llmd2026v09}.

Three current capabilities are especially relevant.

\paragraph{Learned request routing.}
The predicted-latency scheduler trains an online regression model from prompt length, prefix cache hit rate, running requests, queue depth, and KV utilization to estimate TTFT and TPOT on each endpoint \citep{llmd2026latency}. Those features overlap with the inputs an energy predictor would likely need. A natural experiment is to add GPU power/frequency/energy counters and hardware identity and test whether the same routing architecture can estimate incremental joules well enough to influence endpoint selection.

\paragraph{Portable KV reuse.}
Peer-to-peer cache sharing explicitly evaluates whether copying KV state to a better-loaded endpoint beats recomputation \citep{llmd2026p2p}. The synthesis proposed here is to make that crossover function richer: compare not only latency, but also transfer energy, recompute energy, memory pressure, and future reuse probability.

\paragraph{Request-level observability.}
End-to-end OpenTelemetry tracing now spans Gateway, Endpoint Picker, KV-cache, prefill/decode proxy, and model server, exposing the causal path of a request \citep{llmd2026trace}. That tracing path provides a practical place to test request-level attribution by attaching measured or estimated energy and carbon to the same trace that already explains cache score, phase placement, and queue time.

These properties make \texttt{llm-d} more suitable for sustainability orchestration than a model-server-only solution. The sustainability controller needs visibility across replicas and freedom to choose among them; it should not be embedded so deeply in a single engine that it cannot compare heterogeneous endpoints or regions.

\section{Proposed Architecture: Sustainable Inference Control Plane}
Based on the patterns above, I sketch an extension of the \texttt{llm-d} decision loop that I call the Sustainable Inference Control Plane (SICP). This section is a design proposal derived from the cited literature, not a description of functionality that is already implemented in \texttt{llm-d}. The design is intentionally incremental: existing performance signals remain useful, while sustainability signals become additional scorers, constraints, and autoscaling inputs.

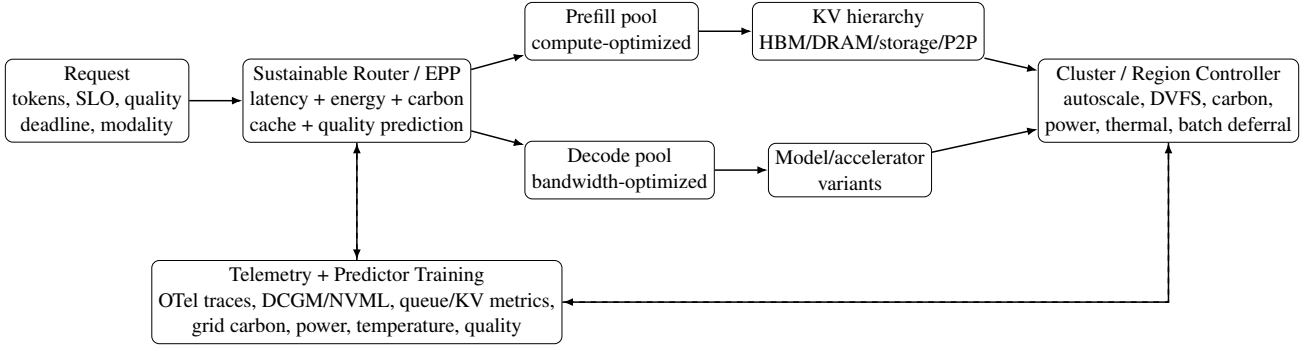
\begin{figure*}[t]
\centering
\resizebox{\textwidth}{!}{%
\begin{tikzpicture}[
node distance=8mm and 10mm,
box/.style={draw,rounded corners,align=center,minimum height=9mm,minimum width=24mm,fill=white},
small/.style={draw,rounded corners,align=center,minimum height=8mm,minimum width=20mm,fill=white},
arrow/.style={-{Latex[length=2mm]},thick}
]
\node[box] (req) {Request\\tokens, SLO, quality\\deadline, modality};
\node[box,right=of req] (router) {Sustainable Router / EPP\\latency + energy + carbon\\cache + quality prediction};
\node[small,right=of router,yshift=13mm] (p1) {Prefill pool\\compute-optimized};
\node[small,right=of router,yshift=-13mm] (d1) {Decode pool\\bandwidth-optimized};
\node[small,right=of p1] (kv) {KV hierarchy\\HBM/DRAM/storage/P2P};
\node[small,right=of d1] (models) {Model/accelerator\\variants};
\node[box,right=of kv,yshift=-13mm] (regions) {Cluster / Region Controller\\autoscale, DVFS, carbon,\\power, thermal, batch deferral};
\node[box,below=22mm of router] (tele) {Telemetry + Predictor Training\\OTel traces, DCGM/NVML, queue/KV metrics,\\grid carbon, power, temperature, quality};

\draw[arrow] (req) -- (router);
\draw[arrow] (router) -- (p1);
\draw[arrow] (router) -- (d1);
\draw[arrow] (p1) -- (kv);
\draw[arrow] (d1) -- (models);
\draw[arrow] (kv) -- (regions);
\draw[arrow] (models) -- (regions);
\draw[arrow] (tele) -- (router);
\draw[arrow] (tele.east) -| (regions.south);
\draw[arrow,dashed] (regions.south) |- (tele.east);
\draw[arrow,dashed] (router.south) -- (tele.north);
\end{tikzpicture}%
}
\caption{Proposed Sustainable Inference Control Plane. Fast request routing predicts SLO feasibility, cache reuse, energy, and carbon; slower controllers change replica counts, phase allocation, power states, model variants, and regional placement. Telemetry closes the loop.}
\label{fig:sicp}
\end{figure*}

\subsection{Fast-path request decision}
For each request $r$ and feasible endpoint/plan $e$, the controller estimates:
\begin{align}
\hat{L}_{r,e} &= f_L(x_r, s_e),\\
\hat{E}_{r,e} &= f_E(x_r, s_e),\\
\hat{C}_{r,e} &= \hat{E}_{r,e}\cdot CI_{g(e)}(t) + C^{emb}_{r,e},\\
\hat{Q}_{r,e} &= f_Q(x_r,m_e),\\
\hat{K}_{r,e} &= f_K(x_r,s_e,m_e),
\end{align}
where $x_r$ includes request properties (input tokens, expected output tokens, modality, reasoning budget, deadline, reusable-prefix fingerprint), $s_e$ includes endpoint state (queue, cache hit, KV pressure, GPU utilization, frequency, hardware, batch state), $CI_{g(e)}(t)$ is time-varying grid carbon intensity for the endpoint region, $C^{emb}_{r,e}$ is an amortized embodied-carbon term when lifecycle-aware accounting is enabled, and $\hat{K}_{r,e}$ is the predicted monetary serving cost of assigning request $r$ to endpoint or execution plan $e$.

The route is selected from the feasible set $\mathcal{F}$ that satisfies quality and SLO constraints:
\begin{equation}
e^* = \arg\min_{e\in\mathcal{F}} \left[
\alpha \hat{L}_{r,e} + \beta \hat{E}_{r,e} + \gamma \hat{C}_{r,e} + \delta \hat{K}_{r,e} + \eta R_{r,e}
\right],
\end{equation}
subject to
\begin{equation}
\hat{TTFT}_{r,e}\leq SLO_{TTFT},\quad
\hat{TPOT}_{r,e}\leq SLO_{TPOT},\quad
\hat{Q}_{r,e}\geq Q_{min}.
\end{equation}
$R_{r,e}$ captures reliability or recomputation risk. The weights need not be static: a datacenter under a demand-response event can increase the power/carbon term, while a latency-critical interactive request can emphasize SLO headroom.

A practical implementation should avoid hiding every trade-off inside one opaque scalar. A simpler starting policy is lexicographic: (1) reject infeasible quality/SLO plans; (2) satisfy reliability and capacity constraints; (3) select from the remaining Pareto frontier using the operator's sustainability policy. This prevents a low-carbon route from silently violating user-visible performance.

\subsection{Cache action as an explicit decision}
When a useful prefix exists elsewhere, the router chooses among four actions: local hit, route-to-owner, peer transfer, or recompute. For a candidate transfer from endpoint $i$ to $j$, transfer is sustainability-preferred when
\begin{equation}
E^{net}_{i\rightarrow j} + E^{load}_{j} + \lambda M^{pressure}_{j}
< E^{recompute}_{j},
\end{equation}
subject to the latency crossover and expected reuse. For agentic workloads, a fifth action---retain for TTL---competes with eviction/offload, following the insight of Continuum \citep{li2025continuum}. This makes cache policy a measurable energy optimization instead of an implicit side effect of routing.

\subsection{Phase-aware power and hardware control}
The phase controller assigns prefill and decode to hardware classes and power states based on phase sensitivity. Splitwise motivates heterogeneous phase placement \citep{patel2024splitwise}; PowerSlider shows why power caps should be allocated differently across phases \citep{li2026powerslider}. In SICP, the autoscaler therefore reasons in units of \emph{workload variants}: e.g., long-context prefill on compute-dense accelerators, decode on memory-bandwidth-efficient devices, or mixed/aggregated execution when network transfer cost outweighs disaggregation benefit.

\subsection{Model and reasoning routing}
The request classifier can additionally choose a model or reasoning mode. Model-routing research shows that many requests do not require the most expensive model \citep{chen2023frugalgpt,ding2024hybrid,ong2024routellm}. The proposal here is to extend that idea to an explicit quality-constrained sustainability decision. Importantly, the controller should measure the energy of both the router and the selected model; a complex routing model that saves few large-model calls can erase its own benefit.

\subsection{Regional and temporal scheduling}
Interactive traffic has limited geographic and temporal flexibility, but batch/offline traffic may have substantial slack. EcoServe identifies the importance of batch inference \citep{li2025ecoserve}; XWind and carbon-aware routing show that location and real-time electricity mix can be useful control signals \citep{reddy2026xwind,bernhard2026carbonrouting}. Therefore, the slow path can defer or shift eligible work under a deadline while keeping latency-sensitive sessions pinned for cache locality and user experience.

\section{Metrics: From Tokens per Second to Sustainable Goodput}
Performance-only metrics can reward waste. Conversely, energy-only metrics can reward unacceptable latency or low-quality models. For evaluation, it is more informative to report a vector of SLO- and quality-constrained metrics.

Let $G_{SLO}$ be useful output tokens (or completed requests) that meet all required latency and quality constraints over interval $T$. Then report:
\begin{align}
G_E &= \frac{G_{SLO}}{E_{total}} && \text{SLO-goodput per joule},\\
G_C &= \frac{G_{SLO}}{C_{op}+C_{emb}} && \text{SLO-goodput per gCO$_2$e},\\
G_W &= \frac{G_{SLO}}{W_{total}} && \text{SLO-goodput per liter water},\\
G_{\$} &= \frac{G_{SLO}}{Cost_{total}} && \text{SLO-goodput per dollar}.
\end{align}

These metrics should be accompanied by TTFT, TPOT/inter-token latency, P50/P95/P99 end-to-end latency, request success rate, model quality, cache hit rate, input/output token throughput, GPU/accelerator utilization, and power. The Stanford ``intelligence per watt'' direction similarly argues for relating capability to energy rather than energy alone \citep{stanford2025ipw}.

\paragraph{Why a vector rather than one green score?}
Operational carbon, embodied carbon, water, cost, and latency are not interchangeable. A single weighted score embeds value judgments and can hide regressions. A Pareto frontier makes trade-offs explicit and allows the same benchmark to support different operator policies.

\section{Experimental Roadmap}
A later implementation study would need to validate SICP with controlled, reproducible experiments rather than treating the design sketch itself as evidence. The following test plan is intended as a practical starting point.

\subsection{Testbed}
At minimum, use two accelerator classes with different compute/memory/power characteristics and 8--16 serving endpoints. A stronger evaluation includes NVIDIA H100/H200/B200-class hardware plus one alternative accelerator family supported by the serving stack. Deploy vLLM and SGLang variants through \texttt{llm-d}, with Prometheus/OpenTelemetry plus DCGM/NVML-equivalent energy telemetry.

\subsection{Workloads}
Use a workload matrix rather than one synthetic trace:
\begin{itemize}
    \item short-chat: small prompts and moderate decode;
    \item long-context/RAG: long shared prefixes and medium decode;
    \item code: long prompts with short-to-medium output;
    \item reasoning: variable, potentially long generation;
    \item agentic: repeated multi-turn prefixes with tool-call pauses;
    \item batch/offline: deadline-flexible inference;
    \item multimodal: encoder + LLM + generation path where supported.
\end{itemize}
Each workload should sweep offered load from underutilized to saturation and vary prefix sharing.

\subsection{Baselines}
Compare against: round-robin; least-loaded; cache-aware routing; predicted-latency routing; energy-only routing; carbon-only routing; and the proposed joint SICP policy. For phase management, compare aggregated serving, fixed P/D disaggregation, and adaptive aggregation/disaggregation. For KV, compare recompute, owner-affinity, offload, and P2P transfer.

\subsection{Ablations}
Ablate each sustainability signal: remove energy prediction, carbon intensity, cache energy, model routing, DVFS, and embodied-carbon term. This is necessary to distinguish true sustainability gains from a generic load-balancing improvement.

\subsection{Measurement discipline}
Energy should be measured at sufficiently high temporal resolution and attributed to requests or batches, not inferred solely from device TDP. Host energy and network/storage energy should be included where material. Carbon calculations should clearly separate measured energy from modeled grid intensity. Report PUE assumptions, carbon-data source, and whether carbon values are average or marginal. Embodied-carbon accounting should state lifetime and utilization amortization assumptions.

\section{Research Questions and Testable Hypotheses}
The proposed architecture yields concrete systems questions.

\paragraph{RQ1: Can latency prediction be extended to energy prediction without expensive profiling?}
Hypothesis: features already used for predicted-latency routing---prompt length, cache hit, queue depth, running requests, KV utilization---plus hardware/power state are sufficient for useful short-horizon incremental-energy prediction.

\paragraph{RQ2: When is cache transfer greener than recomputation?}
Hypothesis: for long shared prefixes on a high-bandwidth fabric, P2P transfer reduces both TTFT and energy, but the crossover shifts with prefix length, network type, source/destination memory tier, and endpoint utilization.

\paragraph{RQ3: Is adaptive P/D disaggregation more energy-efficient than fixed disaggregation?}
Hypothesis: fixed disaggregation wastes energy under some SLO/load regimes, while an adaptive controller can choose aggregation, chunked prefill, or disaggregation based on TTFT/TPOT headroom and transfer cost.

\paragraph{RQ4: Does carbon-aware routing conflict with cache locality?}
Hypothesis: region shifting can reduce operational carbon for batch and stateless traffic, but aggressive shifting of multi-turn sessions can increase recomputation enough to offset carbon gains. Session pinning plus carbon-aware placement of new sessions should dominate naive global routing.

\paragraph{RQ5: How much sustainability gain comes from model routing versus infrastructure routing?}
Hypothesis: task-aware model selection yields the largest savings on heterogeneous query difficulty, while infrastructure routing dominates when a fixed model must be used for quality/compliance reasons. The best system composes both.

\paragraph{RQ6: Can the same control plane manage agentic and multimodal workloads?}
Hypothesis: a graph/phase representation inspired by composite serving systems such as M* \citep{jha2026mstar} can generalize sustainability decisions beyond the two-stage text P/D abstraction.

\section{Future of \texttt{llm-d}: Toward a Resource-Intelligence Control Plane}
Recent \texttt{llm-d} evolution suggests that its long-term value is not a specific optimization but a common decision layer above rapidly changing inference engines. The project has already expanded cache management, heterogeneous engine support, multimodal serving, batch flow control, agentic/RL workflows, GPU-aware routing, and deep tracing \citep{llmd2026v08,llmd2026v09,llmd2026trace}. The next logical step is to make \emph{resource intelligence} first-class.

I see seven concrete engineering directions that follow from the synthesis. These are proposed extensions or experiments; unless explicitly cited as an existing capability, they should not be read as current 	exttt{llm-d} features.

\begin{enumerate}
    \item \textbf{Energy-aware Endpoint Picker plugins.} Add predicted incremental joules and power headroom alongside cache and latency scores.
    \item \textbf{Carbon-aware WorkloadVariantAutoscaling.} Autoscale not only on saturation/SLOs but on carbon intensity and demand-response signals, with hard guardrails for availability.
    \item \textbf{Phase-specific DVFS and accelerator selection.} Map compute-heavy prefill and memory-heavy decode to their energy-efficient hardware/power points.
    \item \textbf{Energy-priced KV orchestration.} Make local hit, owner routing, P2P copy, DRAM/storage offload, and recompute explicit alternatives in one cost model.
    \item \textbf{Quality-aware model/adapter routing.} Integrate small/large model cascades and reasoning-budget selection while preserving a declared minimum quality.
    \item \textbf{Sustainability traces.} Attach estimated/measured joules, grid intensity, cache-reuse avoided compute, and phase placement to the same OpenTelemetry request trace.
    \item \textbf{Cross-cluster federation for flexible work.} Route deadline-tolerant batch inference across regions or renewable-powered sites while protecting cache-local interactive sessions.
\end{enumerate}

If implemented carefully, this would shift the optimization target from ``tokens/second per GPU'' to ``quality- and SLO-satisfied useful work per constrained resource.'' That is a more durable abstraction because accelerators, model architectures, and energy systems will continue to change.

\section{Limitations and Responsible Claims}
The claim boundary in this paper is intentionally conservative. Numerical improvements and measured behaviors cited throughout belong to the original systems and authors, under their hardware, workloads, and measurement assumptions. This paper does not re-measure those results, does not claim that SICP has been implemented, and does not claim that the reported gains would carry over unchanged to an \texttt{llm-d}-based implementation. The contribution is the comparison and connection of those findings into a control-plane design hypothesis. The cited improvements should not be combined arithmetically or treated as additive. For example, a 50\% energy reduction from one optimization and a 40\% carbon reduction from another do not imply a 70\%+ combined reduction. Interactions can be antagonistic: speculative decoding may improve latency but increase energy at high batch; more GPUs can reduce latency while increasing total energy; cache retention can avoid recompute but increase memory pressure; and geographic carbon routing can degrade cache locality.

Carbon accounting also depends on location, time, PUE, and whether average or marginal grid emissions are used. Water impact is even more site-specific and may be dominated by cooling or electricity-generation assumptions. Embodied-carbon accounting depends on hardware lifetime and utilization. These uncertainties argue for reporting raw energy and system behavior alongside modeled environmental metrics.

Finally, sustainability must not become a pretext for quality degradation that is invisible to users. Model routing and reasoning-budget control should be bounded by explicit quality requirements and evaluated on application-relevant tasks.

\section{Conclusion}
The papers reviewed here attack different parts of the LLM serving stack, but they repeatedly expose the same kinds of control variables: workload shape, phase behavior, cache locality, hardware choice, power state, SLO headroom, and placement. My reading of that common pattern is that sustainability is not only a model- or kernel-level optimization problem; it is also an orchestration problem.

That is why \texttt{llm-d} is an interesting place to connect these ideas. A sustainability-aware experiment can build on existing serving engines rather than replace them, and can test whether energy, carbon, cache-recompute cost, and power headroom are useful signals alongside TTFT, TPOT, queue depth, and cache locality.

The paper's main proposal is deliberately testable: treat the path of an inference request as an optimization target, while holding quality and SLOs constant. The next step is implementation and measurement. Until that work is done, the SICP architecture should be read as an engineering hypothesis derived from prior research, not as a claim of measured improvement.

\section*{Acknowledgments}
This manuscript is a research synthesis and systems proposal. All empirical results, numerical improvements, and existing system capabilities discussed in the paper are credited to the cited authors and projects; no new experimental measurements are claimed here. I am grateful to the research and open-source communities whose papers, systems, benchmarks, and technical documentation make this kind of cross-layer comparison possible.

\bibliographystyle{plainnat}
\bibliography{references}

\end{document}